\documentclass[preprint,longbibliography,showpacs,preprintnumbers,amssymb,superscriptaddress,aps,prd,nofootinbib,11pt]{revtex4-1}

\selectfont

\usepackage{graphicx}       % Include figure files
\usepackage{dcolumn}        % Align table columns on decimal point
\usepackage{bm}             % bold math
\usepackage{psfrag}
\usepackage{tensor}
\usepackage[usenames]{color}
\usepackage{amsmath}
\usepackage{subfigure}

\newcommand{\be}{\begin{equation}}
\newcommand{\ee}{\end{equation}}

\newcommand{\etal}{\emph{et al.}}

\newcommand{\beq}{\begin{equation}}
\newcommand{\eeq}{\end{equation}}

\newcommand{\cR}{\mathcal{R}}

\usepackage[]{hyperref}
\hypersetup
	{
		colorlinks,%
		citecolor=blue,%
		linkcolor=red,%
		urlcolor=blue,%
	}

\begin{document}

\preprint{}

\title{Conversion of electromagnetic and gravitational waves by a charged black hole}

\author{Mohamed Ould El Hadj}
 \email{med.ouldelhadj@gmail.com}
\affiliation{Consortium for Fundamental Physics,
School of Mathematics and Statistics,
University of Sheffield, Hicks Building, Hounsfield Road, Sheffield S3 7RH, United Kingdom.}

\author{Sam R. Dolan}
 \email{s.dolan@sheffield.ac.uk}
\affiliation{Consortium for Fundamental Physics,
School of Mathematics and Statistics,
University of Sheffield, Hicks Building, Hounsfield Road, Sheffield S3 7RH, United Kingdom.}

\date{\today}

\begin{abstract}
In a strong electromagnetic field, gravitational waves are converted into electromagnetic waves of the same frequency, and vice versa. Here we calculate the scattering and conversion cross sections for a planar wave impinging upon a Reissner-Nordstr\"om black hole in vacuum, using the partial-wave expansion and numerical methods. We show that, at long wavelengths, the conversion cross section matches that computed by Feynman-diagram techniques. At short wavelengths, the essential features are captured by a geometric-optics approximation. We demonstrate that the converted flux can exceed the scattered flux at large scattering angles, for highly-charged black holes. In the short-wavelength regime, the conversion effect may be understood in terms of a phase that accumulates along a ray. We compute the scattering angle for which the converted and scattered fluxes are equal, as a function of charge-to-mass ratio. We show that this scattering angle approaches $90^\circ$ in the extremal limit.
\end{abstract}



\maketitle

%\tableofcontents

\section{Introduction}

The Gertsenshte\u{\i}n-Zel'dovich (GZ) effect \cite{gertsenshtein1962wave,zel1973electromagnetic} is the conversion of electromagnetic waves into gravitational waves, and vice versa, in the presence of a strong magnetic field. It is a classical (i.e.~non-quantum) phenomenon that is nevertheless extremely weak, since it involves coupling to gravity. In a uniform transverse magnetic field $B_\perp$, electromagnetic waves (EWs) are converted into gravitational waves (GWs), and vice versa, over a length scale of $L = \frac{\pi}{2} \frac{c}{\sqrt{4\pi \epsilon_0 G} \, B_{\perp} } \approx 1.77 \, \text{Mpc} \left( \frac{B_{\perp}}{1 \, \text{Gauss}} \right)^{-1}$. The effect is potentially significant in the early universe \cite{Dolgov:2012be,Fujita:2020rdx}, where the combination of cosmic magnetic fields and primordial gravitational waves could generate distortions of the CMB spectrum \cite{Domcke:2020yzq}.

The interconversion of EWs and GWs in various non-uniform electromagnetic field configurations has been studied from a theoretical perspective. In 1977, De Logi and Mickelson \cite{DeLogi:1977qe} applied Feynman perturbation techniques to study ``catalytic'' conversion in static electromagnetic fields. The low-energy graviton-to-photon ($g\rightarrow\gamma$) conversion cross section in the Coulomb field of a fixed charge $Q$ in SI units is \cite{DeLogi:1977qe, Bjerrum-Bohr:2014lea}
\beq
\frac{d\sigma}{d\Omega}^{g \rightarrow \gamma} = \frac{G Q^2}{4 \pi \epsilon_0 c^4} \cot^2(\theta / 2) \left( \cos^4(\theta/2) + \sin^4(\theta/2) \right) . \label{eq:delogi}
\eeq
The first (second) term in parentheses is associated with the cross section for generating an electromagnetic wave of the same (opposite) handedness as the incident gravitational wave \cite{DeLogi:1977qe}. The conversion cross section exhibits a $\theta^{-2}$ divergence in the forward direction, distinct in character from the more familiar Rutherford divergence ($\theta^{-4}$).
The dominant contribution is from the photon-pole (`t-pole') Feynman diagram, rather than the `seagull' diagram of Compton scattering \cite{Bjerrum-Bohr:2014lea}.

A macroscopic realization of a fixed charge is the Reissner-Nordstr\"om (RN) black hole, of mass $M$ and charge $Q$. In classical field theory, a RN black hole can support charges of up to $Q_{\text{max}} = M \sqrt{4 \pi \epsilon_0 G} \approx 1.7 \times 10^{20} (M/M_\odot) \, \text{C}$. On the other hand, astrophysical black holes are unlikely to possess a charge any greater than 1C per solar mass, some twenty orders of magntiude lower, due to charge-neutralization effects \cite{Gibbons:1975kk}.

The conversion of EW and GWs by a RN black hole has been addressed by several authors since 1974 \cite{Gerlach:1974zz, Moncrief-1974a,Moncrief-1974b,Moncrief-1975,Olson:1974nk,Matzner:1976kj,Chandrasekhar:1979iz,Gunter:1980,Breuer:1981kd}. Gerlach \cite{Gerlach:1974zz} showed that, in the high-frequency (geometric-optics) regime, there is a beating between electromagnetic and gravitational modes, associated with a periodic transfer of energy. Moncrief \cite{Moncrief-1974a,Moncrief-1974b,Moncrief-1975} reduced the coupled system of electromagnetic and gravitational perturbation equations on the RN spacetime to a \emph{pair} of decoupled second-order ODEs, for each parity and angular harmonic. Olson and Unruh \cite{Olson:1974nk} studied the odd-parity sector in the WKB regime. Matzner \cite{Matzner:1976kj} addressed the conversion of a incident planar wave via the partial-wave approach, focussing particularly on the $\ell = 2$ mode. Fabbri \cite{Fabbri:1977} also addressed scattering and conversion cross sections. Breuer \etal~\cite{Breuer:1981kd} calculated the conversion scattering cross section under the Born approximation. Chandrasekhar \cite{Chandrasekhar:1979iz} extended the work of Moncrief, clarifying the phase relationship between odd and even parity perturbations. Gunter \cite{Gunter:1980} calculated phase shifts, conversion factors and quasi-normal mode frequencies.  Torres del Castillo \cite{Castillo:1987,Castillo:1996jm} derived asymptotic expressions for Maxwell and Weyl scalars from Hertz-Debye potentials.  More recently, the $\gamma \rightarrow \gamma$ and $g \rightarrow g$ scattering cross sections were calculated by Crispino \etal~in Refs.~\cite{Crispino:2014eea,Crispino:2015gua}, and scalar-field scattering was examined in Ref.~\cite{Crispino:2009ki}. The construction of the metric perturbation and vector potential in Regge-Wheeler gauge was described in Refs.~\cite{Zhu:2018tzi, Burton:2020wnj}.

In this work, we calculate the scattering and conversion cross sections for a monochromatic planar wave impinging upon a RN black hole, using both the partial-wave method and a geometric-optics approximation. The scattering scenario is described by a pair of dimensionless parameters, $M \omega$ and $Q/M$. We focus particularly on the differential cross section for the conversion of an incident electromagnetic wave to an outgoing gravitational wave, which is equal to the cross section for the opposite process.

The article is organised as follows. After introducing the linearized Einstein-Maxwell system in Sec.~\ref{sec:EinMax}, we review the separation of variables method achieved by Moncrief for the RN black hole in Sec.~\ref{sec:perturbations}. The partial-wave expressions for the scattering amplitudes and cross sections are summarized in Sec.~\ref{sec:summary}. Notes on the numerical method in Sec.~\ref{sec:numerical-method} are followed by a description of the geometric optics approximation in Sec.~\ref{sec:geo-optics}. The key results are presented in Sec.~\ref{sec:results}, and we conclude with a discussion in Sec.~\ref{sec:conclusions}. Throughout we adopt units such that $G = c = 4 \pi \epsilon_0 = 1$, and $\nabla_\mu$ denotes the covariant derivative.

\begin{figure}%[htb]
 \includegraphics[scale=0.65]{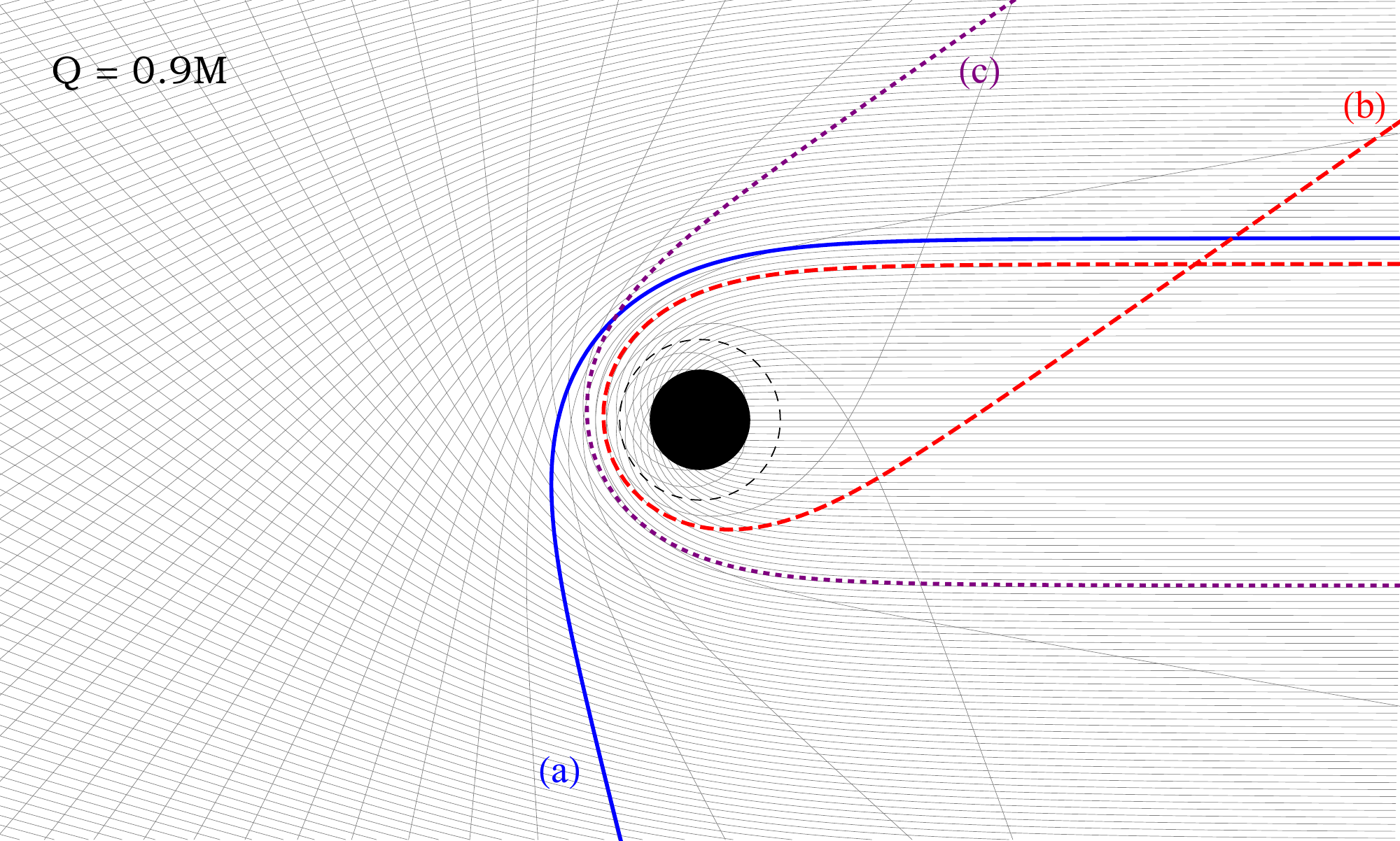}
 \caption{\label{Fig:ray-plot}
 Parallel rays impinging upon a Reissner-Nordstr\"om black hole from the right.
 In the geometric-optics approximation (high-frequency/short-wavelength limit), the part of the wave incident along ray (a) is half-converted from gravitational waves to electromagnetic waves (or vice versa), and the part of the wave along ray (b) is totally converted. Ray (c) emerges at the same angle as ray (b), generating an interference effect (`orbiting') in the cross sections. The dashed black circle shows the photon orbit.
  }
\end{figure}

\section{Method}

\subsection{The linearized Einstein-Maxwell system\label{sec:EinMax}}
The equations of the linearized Einstein-Maxwell system take the form
\begin{subequations}
\begin{align}
\Box \overline{h}_{\mu \nu} &= - \frac{16 \pi G}{c^4} \delta T_{\mu \nu} , \\
\Box \alpha_{\mu} &= \delta S_\mu ,
\end{align}
\label{eq:linearized}
\end{subequations}
where $\overline{h}_{\mu \nu}$ is the trace-reversed metric perturbation and $\alpha_\mu$ is the perturbation in the vector potential, in Lorenz gauge \cite{Gerlach:1974zz}. The source term $\delta S_\mu$ in the electromagnetic field equation is
\beq
\delta S_\mu \equiv 2 g^{\nu \gamma} \delta \tensor{\Gamma}{^\sigma _{\nu [\mu}} F_{\gamma] \sigma}
\eeq
where $\tensor{\delta \Gamma}{^\sigma _{\nu \mu}}$ is the perturbation in the Christoffel connection due to the metric perturbation, and $g^{\nu \gamma}$ and $F_{\gamma \sigma}$ are the background inverse metric, and background electromagnetic field tensor, respectively. In these expressions, the covariant derivative is taken with respect to the unperturbed spacetime.
The source term in the gravitational field equation $\delta T_{\mu \nu}$ is given in (e.g.) Eq.~(4) of Ref.~\cite{Gerlach:1974zz}. In the following sections, we address the linearized Einstein-Maxwell system for the specific case in which the background spacetime and field tensor correspond to a charged black hole.

\subsection{Perturbations of the Reissner-Nordstr\"om spacetime\label{sec:perturbations}}
A RN black hole of mass $M$ and charge $Q$ is described by the line element
\beq
ds^2 = -f(r) dt^2 + f^{-1}(r) dr^2 + r^2 d\Omega^2,
\eeq
where $f(r) = 1 - 2M / r + Q^2/r^2 =  (1-r_+/r)(1-r_-/r)$ with $r_\pm = M \pm \sqrt{M^2 -Q^2}$, and $d\Omega^2 = d\theta^2 +\sin^2\theta \, d\phi^2$ denotes the line element on the unit $2$-sphere $S^2$.

Moncrief \cite{Moncrief-1974a,Moncrief-1974b,Moncrief-1975} showed that the set of coupled equations governing the metric perturbation and vector potential is amenable to a separation of variables. Assuming harmonic time dependence ($e^{-i \omega t}$), and separating with spin-weighted spherical harmonics, ${}_sY_\ell^m(\theta)$, the dynamical degrees of freedom are encapsulated by a pair of radial functions $\phi_{1 \omega\ell}^{(e/o)}$ and  $\phi_{2 \omega\ell}^{(e/o)}$ for each parity ($e/o$), which are governed by second-order ordinary differential equations, viz.,
\beq
\left\{\frac{d^2}{dr_{*}^2}+\omega^2- V_{s \ell}^{(e/o)}(r) \right\} \phi_{s \omega\ell}^{(e/o)}(r) = 0 .
\eeq
Here $s \in \{1, 2\}$, and the symbol $e$ ($o$) denotes even (odd) parity. The tortoise coordinate $r_\ast$ is defined by $dr_\ast / dr = 1/f(r)$.

The odd-parity potential is
\beq
V_{s \ell}^{(o)}(r) = f(r) \left[ \frac{\Lambda + 2}{r^2} - \frac{q_{s}}{r^3} \left(1 + \frac{6M -q_{s}}{\Lambda r}\right)  \right],
\eeq
where $\Lambda\equiv (\ell - 1)(\ell + 2)$ and
\begin{align}
q_1 &= 3M - \sqrt{9M^2 + 4 \Lambda Q^2} , \\
q_2 &= 3M + \sqrt{9M^2 + 4 \Lambda Q^2} .
\end{align}
(N.B.~Here $q_1$ and $q_2$ are defined in the opposite order to in Refs.~\cite{Chandrasekhar:1979iz,Gunter:1980}, in order to simplify the subsequent expressions.)

Chandrasekhar \cite{Chandrasekhar:1979iz,Chandrasekhar:1985kt} showed that Moncrief's even-parity potential can be written as
\beq
V_{s \ell}^{(e)}(r) = V_{s \ell}^{(o)}(r) + 2 q_{s} \frac{d}{dr_\ast} \left[ \frac{f(r)}{r (q_{s} + \Lambda r)}  \right],
\eeq
and odd and even-parity functions are related by \cite{Chandrasekhar:1979iz,Chandrasekhar:1985kt, Gunter:1980}
\begin{equation}\label{eq:odd-even-relation}
  \left[\vphantom{\frac{d}{dr}}\Lambda(\Lambda+2) \mp 2 i \omega q_{s}\right]\phi_{s \omega\ell}^{\text{(e/o)}} = \left[\Lambda(\Lambda+2)+\frac{2q_{s}^2}{r\left(\Lambda r+q_{s}\right)}f(r)\pm 2\,q_{s}f(r)\frac{d}{dr}\right]\phi_{s \omega\ell}^{\text{(o/e)}}.
\end{equation}
Here, the upper (lower) sign is associated with the first (second) choice of parity in the superscript.

We now define the modes that are ingoing at the horizon (i.e.~in the limit $r \rightarrow r_+$ and $r_\ast \rightarrow -\infty$),
\beq
\phi_{s \omega\ell}^{\text{in} (e/o)}(r) =
\begin{cases}
   e^{-i \omega r_\ast} , & r_\ast \rightarrow - \infty , \\
 A^{(-,e/o)}_{s \ell\omega}\, e^{-i \omega r_\ast} +  A^{(+,e/o)}_{s \ell\omega}\, e^{+ i \omega r_\ast} , &  r_\ast \rightarrow + \infty ,
\end{cases}
\label{eq:Znorm}
\eeq
where the coefficients $ A^{(\pm,e/o)}_{s \ell\omega}$ are complex amplitudes such that $\big{|}A^{(-,e/o)}_{s \ell\omega}\big{|}^2 - \big{|}A^{(+,e/o)}_{s \ell\omega} \big{|}^2 = 1$. It follows from Eq.~(\ref{eq:odd-even-relation}) that the coefficient $ A^{(-,e/o)}_{s \ell\omega}$ does not depend on parity
\begin{equation}\label{Rel_1_Am}
A^{(-,e)}_{s \ell\omega} = A^{(-,o)}_{s \ell\omega} \equiv  A^{(-)}_{s \ell\omega},
\end{equation}
whereas
\begin{equation}\label{Rel_2_Ap}
 \frac{A^{(+,e)}_{s \ell\omega}}{A^{(+,o)}_{s \ell\omega}} =
              \frac{\Lambda(\Lambda+2)+2i\omega q_{s}}{\Lambda(\Lambda+2)-2i\omega q_{s}} .
\end{equation}
The reflection coefficient $\mathcal{R}_s^{(e/o)}$ is defined by
\begin{align}
  \mathcal{R}_s^{(e/o)} &\equiv \frac{ A^{(+,e/o)}_{s \ell\omega}}{A^{(-)}_{s \ell\omega}} . \label{coef_R}
\end{align}

The electromagnetic ($H$) and gravitational ($Q$) perturbations are derived from the radial functions \cite{Moncrief-1974a,Moncrief-1974b,Moncrief-1975}
\begin{subequations} \label{eq:HQ}
\begin{align}
 H^{(e/o)} &\equiv \phantom{P} \cos \alpha \, \phi_{1 \omega\ell}^{(e/o)} - P \sin \alpha \, \phi_{2 \omega\ell}^{(e/o)} , \\
 Q^{(e/o)} &\equiv P \sin \alpha \, \phi_{1 \omega\ell}^{(e/o)} + \phantom{P} \cos \alpha \, \phi_{2 \omega\ell}^{(e/o)} ,
\end{align}
\end{subequations}
where $P =+1$ for even parity and $P = -1$ for odd parity. Here,
\begin{align}
\cos^2 \alpha &= \frac{q_2}{q_2 - q_1}, &
\sin^2 \alpha &= \frac{-q_1}{q_2 - q_1}, &
\sin(2 \alpha) &= \frac{- 2 \sqrt{-q_1 q_2}}{q_2 - q_1} = \frac{-2 Q \Lambda^{1/2}}{\sqrt{9M^2 + 4 \Lambda Q^2}}.
\label{eq:sincos}
\end{align}
The conversion coefficient $\mathcal{C}^{(e/o)}$ is
\beq
\mathcal{C}^{(e/o)} = \left| \frac{1}{2} \sin(2 \alpha) \left(\cR_1^{(e/o)} - \cR_2^{(e/o)} \right) \right|^2. \label{eq:conv-coeff}
\eeq
This is the fraction of the incident wave (in an $\ell \omega$ mode of parity $(e/o)$) that is converted from electromagnetic to gravitational, and vice versa \cite{Chandrasekhar:1985kt, Crispino:2009zza}.

\subsection{Scattering amplitudes\label{sec:summary}}
The amplitudes for planar wave scattering by a RN black hole are summarized below. A full derivation is given in Ref.~\cite{OuldElHadj:2021} (see also Refs.~\cite{Matzner:1976kj,Futterman:1988ni} for details). Here $\mathfrak{f}$ is the helicity-preserving amplitude, and $\mathfrak{g}$ is the helicity-reversing amplitude. The superscripts $s_i$ and $s_f$ refer to the spins of the initial and final fields, with $s=1$ for an electromagnetic wave and $s=2$ for a gravitational wave. For example, $\mathfrak{f}^{(1 1)}$ is the amplitude for scattering a circular-polarized incoming electromagnetic wave to an outgoing electromagnetic wave of the same handness, and $\mathfrak{g}^{(1 2)}$ is the amplitude for converting an incoming electromagnetic wave ($s_i = 1$) to an outgoing gravitational wave ($s_f = 2$) of the opposite handedness.

The amplitudes are
\begin{subequations}
\label{eq:fg}
\begin{align}
\mathfrak{f}^{(s_i s_f)}(\theta) &= \frac{\pi}{i \omega} \sum_{\ell=\ell_{\text{min}}}^\infty \left[ \left(S_{\ell}^{(e, s_i s_f)} + S_{\ell}^{(o, s_i s_f)}\right) - 2\delta_{s_i s_f} \right] {}_{-s_i} Y_\ell^{s_i}(0)  \, {}_{-s_f} Y_\ell^{s_i}(\theta) ,  \label{eq:f} \\
\mathfrak{g}^{(s_i s_f)}(\theta) &= \frac{\pi}{i \omega} \sum_{\ell=\ell_{\text{min}}}^\infty  (-1)^\ell \left[ \left(S_{\ell}^{(e, s_i s_f)} - S_{\ell}^{(o, s_i s_f)} \right) \right] {}_{-s_i} Y_\ell^{s_i}(0) \, {}_{-s_f} Y_\ell^{s_i}(\pi - \theta) ,  \label{eq:g}
\end{align}
where $\ell_{\text{min}} \equiv \text{max}(s_i, s_f)$ and $\delta_{s_i s_f}$ is the Kronecker delta. Here ${}_{S}Y_\ell^m(\theta)$ are spin-weighted spherical harmonics. The associated cross section is
\beq
\frac{d\sigma}{d\Omega}^{s_i \rightarrow s_f} = \left| \mathfrak{f}^{(s_i s_f)}(\theta) \right|^2 + \left| \mathfrak{g}^{(s_i s_f)}(\theta) \right|^2 . \label{eq:sigma-ee}
\eeq
\end{subequations}
The scattering coefficients in Eqs.~(\ref{eq:fg}) are
\begin{subequations}
\begin{align}
S_{\ell}^{(e/o, 1 1)} &= (-1)^{\ell+1} \left( \cos^2 \alpha \, \cR_1^{(e/o)} + \sin^2 \alpha \, \cR_2^{(e/o)} \right) , \label{eq:See} \\
S_{\ell}^{(e/o, 2 2)} &= (-1)^{\ell+1} \left( \sin^2 \alpha \, \cR_1^{(e/o)} + \cos^2 \alpha \, \cR_2^{(e/o)} \right) , \label{eq:Sgg} \\
S_{\ell}^{(e/o, 1 2)} = S_{\ell}^{(e/o, 2 1)} &= (-1)^{\ell+1} \,\frac{1}{2}\sin(2 \alpha) \left( \cR_1^{(e/o)} - \cR_2^{(e/o)} \right) , \label{eq:Seg}
\end{align}
\end{subequations}
where $\sin^2 \alpha$, $\cos^2 \alpha$ and $\sin(2 \alpha)$ are defined in Eq.~(\ref{eq:sincos}), and the reflection coefficients in Eq.~(\ref{coef_R}). The square magnitude of the scattering coefficient $S_{\ell}^{(e/o, 1 2)}$ is equal to the conversion coefficient $\mathcal{C}^{(e/o)}$ of Eq.~(\ref{eq:conv-coeff}).

In this work, we focus on the conversion cross sections ($s_i \neq s_f$). The $\gamma \rightarrow g$ and $g \rightarrow \gamma$ conversion cross sections are equal, $\frac{d\sigma}{d\Omega}^{1\rightarrow2} = \frac{d\sigma}{d\Omega}^{2\rightarrow1}$. In the Schwarzschild limit ($Q \rightarrow 0$ $\Rightarrow$ $\alpha \rightarrow 0$), the conversion cross sections vanish ($\mathfrak{f}^{(12)} = \mathfrak{g}^{(12)} = 0$). In this same limit, the helicity-reversing amplitude vanishes in the electromagnetic case ($\mathfrak{g}^{(1 1)} = 0$), but not in the gravitational case ($\mathfrak{g}^{(2 2)} \neq 0$).

\subsection{Numerical method\label{sec:numerical-method}}

In order to numerically construct the scattering cross sections~\eqref{eq:sigma-ee}, \textit{i.e.}, the scattering amplitudes~\eqref{eq:f}~and~\eqref{eq:g} for the different processes ($\gamma \to \gamma$, $g \to g$ and $\gamma \to g$), we used the numerical calculation methods of Refs.~\cite{Folacci:2019cmc,Folacci:2019vtt}, which enable the construction of the differential scattering cross sections for scalar, electromagnetic and gravitational waves by a Schwarzschild black hole (see also the extension of these numerical methods to the compact objects in Ref.~\cite{OuldElHadj:2019kji}). Because the scattering amplitudes~\eqref{eq:f}~and~\eqref{eq:g}  suffer of a lack of convergence due to the long range nature of the fields propagating on the RN black hole, we have used the method described in the Appendix of Ref.~\cite{Folacci:2019cmc} to accelerate the convergence of this sum. Finally, all these numerical calculations are performed by using \textit{Mathematica}.

\subsection{Rays and the geometric-optics approximation\label{sec:geo-optics}}
In the short-wavelength (high-frequency) regime, wave propagation is typically well described by a geometric-optics approximation, in which null geodesics (`rays') play a central role. As an example, consider the equation governing a vector potential $\alpha^\mu$ in Lorenz gauge in vacuum,
\beq
\Box \alpha^\mu = 0 , \quad \quad \nabla_{\mu} \alpha^\mu = 0 . \label{eq:Avac}
\eeq
A suitable geometric-optics ansatz is
\beq
\alpha_\mu = \mathcal{A} \, s_\mu e^{i \Phi(x) / \epsilon}, \label{eq:A-geo-optics}
\eeq
where $\mathcal{A}$ is the amplitude, $s_\mu$ is the polarization vector, $\Phi(x)$ is the phase, and $\epsilon=1$ is an order-counting parameter. Inserting (\ref{eq:A-geo-optics}) into (\ref{eq:Avac}) and expanding in powers of $1/\epsilon$ leads to a self-consistent set of equations,
\begin{align}
k_\mu k^\mu &= 0, & k^\mu s_\mu  &= 0 , \\
\nabla_\mu \left( \mathcal{A}^2 k^\mu \right) &=0 & k^\mu \nabla_\mu s_\nu &= 0 ,
\end{align}
where $k_\mu \equiv \nabla_\mu \Phi$. The geodesic equation $k^\mu \nabla_\mu k^\nu = 0$ follows from the null condition $k_\mu k^\mu = 0$ and the fact that $k_\mu$ is a gradient. Moreover, the polarization $s^\mu$ is parallel-transported along the geodesic, and the amplitude $\mathcal{A}$ is inversely proportional to the cross-sectional area of ray bundles.

Gerlach \cite{Gerlach:1974zz} has extended the geometric-optics method to address the problem of the conversion of EM and GWs in a background field, starting with the linearized Einstein-Maxwell equations, Eqs.~(\ref{eq:linearized}). The key result of Ref.~\cite{Gerlach:1974zz} (Eq.~(26)) is that, along a ray, the (normalized) EM and GW amplitudes $\mathcal{A}_\gamma$ and $\mathcal{A}_g$ obey
\begin{subequations} \label{eq:conv-oscillation}
\begin{align}
\mathcal{A}_\gamma = \mathcal{A}(\lambda) \sin \chi(\lambda) , \\
\mathcal{A}_g =  \mathcal{A}(\lambda) \cos \chi(\lambda) ,
\end{align}
\end{subequations}
where $\chi$ is the \emph{conversion phase}, defined by an integral along the ray,
\beq
\chi(\lambda) = -\sqrt{\frac{8 G}{c^4}} \int e_{\mu} \, \hat{e}^{\mu \nu} \, F_{\nu \sigma} \, k^\sigma \, d \lambda.  \label{eq:chi1}
\eeq
In this expression, $e_{\mu}$ and $\hat{e}^{\mu \nu}$ are the (unit) polarization vectors of the EM and GW fields, which are parallel-transported along the ray, and $F_{\nu \sigma}$ is the Faraday tensor for the background field.

For a RN black hole, there is only one non-trivial component of the background field, $F_{tr} = -F_{rt} = Q / r^2$. Inserting the circular polarizations into Eq.~(\ref{eq:chi1}), and setting $G=c=1$,
\beq
\chi(\lambda) = Q \int \frac{ (s^r k^t - s^t k^r) }{r^2} d \lambda . \label{eq:chi-integral}
\eeq

A ray in the equatorial plane ($\theta = \pi/2$) has a tangent vector $k^\alpha \equiv dx^\alpha / d\lambda = [E / f, \dot{r}, 0, E b / r^2]$, where $E$ and $b$ are constants of motion associated with time-translation and axial symmetries of the spacetime. Without loss of generality, we choose an affine parameter $\lambda$ such that $E = 1$, so that $b$ has the interpretation of an impact parameter of the ray. The null condition $g_{\mu \nu} k^\mu k^\nu = 0$ yields the energy equation, $\dot{r}^2 = 1 - f(r) b^2 / r^2 \equiv U(r; b)$. Solving $U(r; b) = \partial_r U(r ; b) = 0$ yields
\begin{align}
r_{c} &= \tfrac{1}{2} M \left( 3 + \sqrt{9 - 8 (Q/M)^2} \right) , &
b_c &= \frac{r_{c}}{\sqrt{f(r_{c})}}  ,  \label{eq:bc}
\end{align}
where $b_c$ is the critical impact parameter for the ray that asymptotes to the photon orbit at radius $r_{\text{c}}$ (shown as a dashed circle in Fig.~\ref{Fig:ray-plot}). On the photon orbit, $dt / d\phi = b_c$. Rays with $b > b_c$ are scattered, and rays with $b < b_c$ are absorbed.

The orbital equation for a ray with impact parameter $b$ is
\beq
\left( \frac{du}{d\phi} \right)^2 = \frac{M^2}{b^2} - u^2 (1 - 2u + (Q/M)^2 u^2)  \equiv \beta(u) ,
\eeq
where $u \equiv M/r$. The deflection angle for a ray, $\Theta$, can be expressed formally as an integral,
\beq
\Theta = 2 \int_0^{u_1} \frac{du}{\sqrt{\beta (u)}} - \pi , \label{eq:deflection-integral}
\eeq
where $u_1$ is the first positive root of the polynomial $\beta(u)$. 

To determine the conversion phase $\chi$ in a similar form, we note that expression (\ref{eq:chi-integral}) is invariant under $s^\mu \rightarrow \tilde{s}^\mu = s^\mu + \kappa(\lambda) k^\mu$, where $\kappa(\lambda)$ is an arbitrary function. Hence it is not necessary to solve the parallel-transport differential equation directly. Instead, we may simply insert a spatial unit spin vector of the form $\tilde{s}_\mu = [0, \tilde{s}_r, 0, \tilde{s}_\phi]$ and impose the algebraic constraints $g^{\mu\nu} \tilde{s}_\mu \tilde{s}_\nu = 1$ and $\tilde{s}_\mu k^\nu = 0$ to determine that $\tilde{s}_r = b / r$. Consequently,
\beq
\chi = \frac{2Q}{M} \int_0^{u_1} \frac{u du}{\sqrt{\beta(u)}}. \label{eq:conv-integral}
\eeq
In general, Eq.~(\ref{eq:deflection-integral}) and Eq.~(\ref{eq:conv-integral}) may be expressed in terms of elliptic integrals. In the special case $Q=M$ a certain linear combination of (\ref{eq:deflection-integral}) and (\ref{eq:conv-integral}) has an elementary solution, viz.,
\beq
2 \chi - \Theta = \pi + 2 \int_0^{u_1} \frac{(2 u  - 1) du}{\sqrt{\frac{M^2}{b^2} - \left(u (1 - u) \right)^2}}  = 0 ,
\eeq
where $u_1 (1 - u_1) = M / b$, and the integral is performed with the substitution $v = u(1-u)$. 
Thus for an extremal black hole ($Q=M$) there is a straightforward linear relationship between the deflection angle and the conversion phase,
\beq
\chi = \frac{1}{2} \Theta .  \label{eq:chi-linear} 
\eeq

The classical scattering cross section is the ratio of the area on on the initial wavefront of a family of rays, $\delta A = 2 \pi b \, \delta b$, to the solid angle into which they are scattered, $\delta \Omega = 2 \pi \sin \theta |d \Theta|$, that is,
\beq
\left. \frac{d\sigma}{d\Omega} \right|_{cl.}  = \frac{b}{ \sin \theta \left| d\Theta / db \right|} , \label{eq:classical}
\eeq
where $\Theta(b)$ is the deflection function. Implicit in Eq.~(\ref{eq:classical}), however, are the assumptions that $\Theta(b)$ is an invertible function (i.e.~that there is a single ray associated with a given scattering angle), and that there is no conversion. Below we seek an extended approximation that remedies both deficiencies. (The classical cross section also fails where the denominator vanishes, i.e.~for glories ($\theta = n \pi$) and for rainbows ($\Theta' = 0$). To handle these cases, a more sophisticated semi-classical analysis is required, but this is not pursued here.)

To obtain a (numerical) geometric-optics approximation to the scattering and conversion cross sections, we took the following steps:
\begin{enumerate}
 \item Define a time function $T(b; x_i, r_f)$ (with $x_i , r_f \gg M$ fixed parameters, with $x_i, r_f \sim 1000M$ sufficient for our purposes) corresponding to the coordinate time that it takes for a ray starting on a planar wave front, a perpendicular distance $x_i$ from the origin, to reach to a radius $r = r_f$ after scattering (N.B.~$T(b)$ diverges as $b \rightarrow b_c$ and is undefined for $b < b_c$).
 \item Define the deflection function $\Theta(b; x_i, r_f)$ in a similar way, from the $\phi$ coordinate at $r=r_f$. 
 \item Solve the parallel-transport equation $k^\mu \nabla_\mu s_\nu = 0$ to calculate the $t$ and $r$ components of the spin vector, starting with initial conditions such that $s_\mu k^\nu = 0$ and $g^{\mu \nu} s_\mu s_\nu = 1$ (relations which are preserved along the ray).
 \item Calculate the conversion phase $\chi(b; x_i, r_f)$ from the integral along the ray using Eq.~(\ref{eq:chi-integral}).
 \item For a given scattering angle $\theta$, define pseudo-amplitudes
 \begin{subequations} \label{eq:geo-ampl}
 \begin{align}
f_{\text{geo}}^{(\text{scat})} &= \sum_{k=1}^\infty \mathcal{A}_k e^{- i \omega T(b_k)} (-1)^k \cos \chi_k , \\
f_{\text{geo}}^{(\text{conv})} &= \sum_{k=1}^\infty \mathcal{A}_k e^{- i \omega T(b_k)} \sin \chi_k ,
 \end{align}
 \label{eq:geo-optics-ampl}
  where
  \beq
  \mathcal{A}_k = \sqrt{ \frac{b_k}{\sin(\theta) \left| \frac{d \Theta_k}{d b} \right|}} .
  \eeq
  \end{subequations}
  \item The geometric-optics approximations for the scattering and conversion cross sections are given by the square-magnitudes of the amplitudes,
  \begin{subequations} \label{eq:geo-csec}
  \begin{align}
  \frac{d\sigma}{d\Omega}^{\gamma \rightarrow \gamma}_{\text{geo}} = \frac{d\sigma}{d\Omega}^{g \rightarrow g}_{\text{geo}} &= \left| f_{\text{geo}}^{(\text{scat})}  \right|^2 , \\
  \frac{d\sigma}{d\Omega}^{\gamma \rightarrow g}_{\text{geo}} = \frac{d\sigma}{d\Omega}^{g \rightarrow \gamma}_{\text{geo}} &= \left| f_{\text{geo}}^{(\text{conv})}  \right|^2 .
  \end{align}
  \end{subequations}
  \end{enumerate}
 In principle, the sum in Eq.~(\ref{eq:geo-optics-ampl}) is taken over \emph{all} rays that emerge at angle $\theta$, in other words, rays with deflection angles $\Theta_1 = \theta$, $\Theta_2 = 2 \pi - \theta$, $\Theta_3 = 2 \pi + \theta$, $\ldots$ and corresponding impact parameters $b_1$, $b_2$, $b_3$ etc. In practice, we sum the contributions from only the primary ($k=1$) and secondary ($k=2$) rays, as this is sufficient to reproduce the orbiting oscillation visible in the results. 

The divergences in the cross sections in the small-angle limit ($\theta \rightarrow 0$) may be understood in terms of rays in the weak field ($b \gg M$). Using the Einstein deflection angle, $\Theta \sim 4M/b \ll 1$, and the conversion phase for a ray in Minkowski spacetime, $\chi \sim 2 Q / b \ll 1$, the scattering and conversion cross sections scale as
  \begin{subequations} \label{eq:geo-csec-divergence}
  \begin{align}
  \frac{d\sigma}{d\Omega}^{\gamma \rightarrow \gamma}_{\text{geo}} = \frac{d\sigma}{d\Omega}^{g \rightarrow g}_{\text{geo}} &\sim  \frac{16M^2}{\theta^4} , \\
  \frac{d\sigma}{d\Omega}^{\gamma \rightarrow g}_{\text{geo}} = \frac{d\sigma}{d\Omega}^{g \rightarrow \gamma}_{\text{geo}} &\sim  \frac{4 Q^2}{\theta^2} ,
  \end{align}
  \end{subequations}
  at small angles.

\section{Results\label{sec:results}}

\begin{figure}%[htb]
 \includegraphics[scale=0.70]{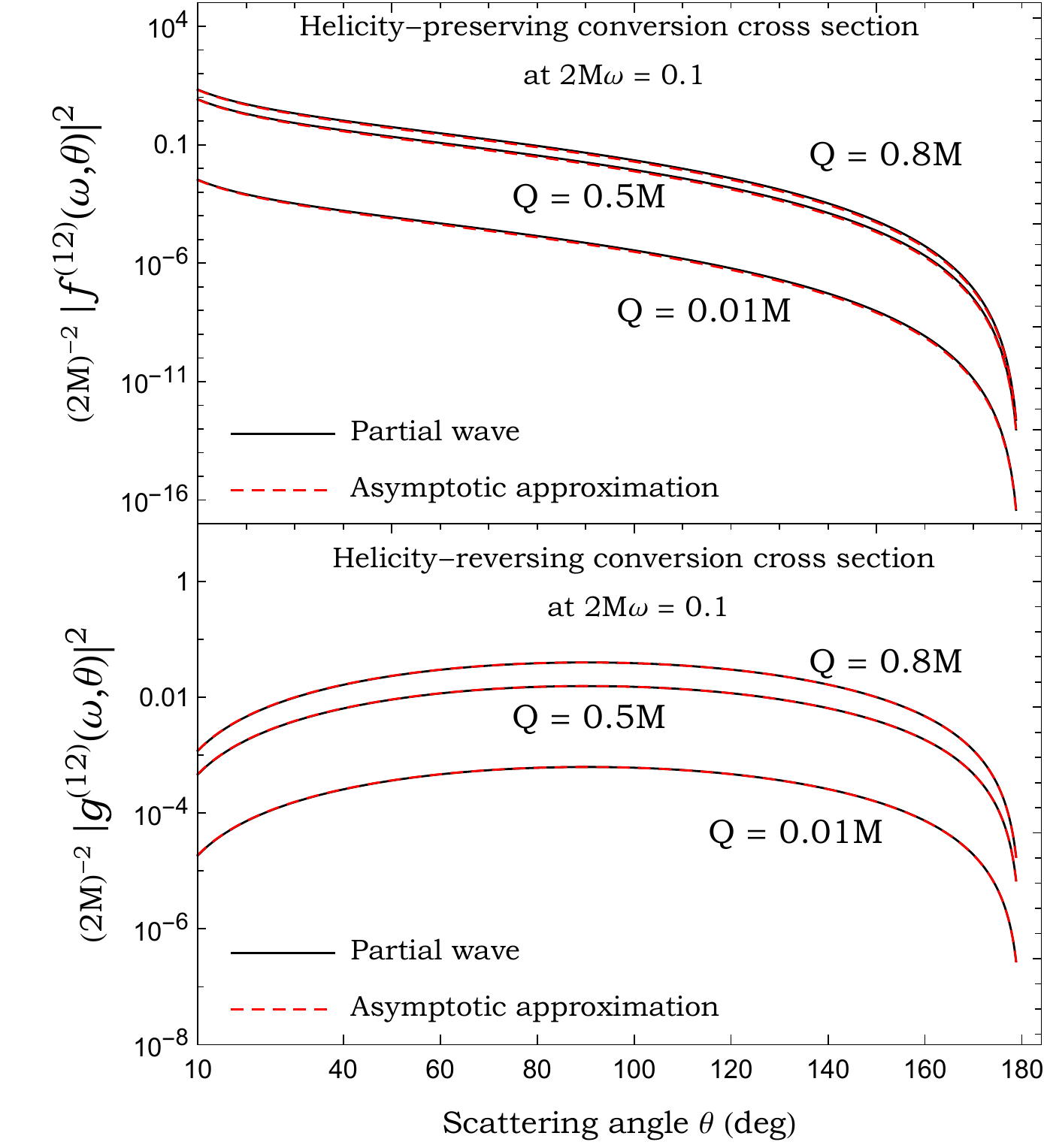}
 \caption{\label{Fig:fp_fm_Conv_Exact_Approx_Diff_Q} Comparison of helicity-preserving and helicity-reversing conversion scattering amplitude with asymptotic approximation at low frequency ($2M\omega=0.1$) for $Q=0.01M, 0.5M$ and $0.8M$. The solid line shows the cross section computed from the partial-wave series, Eq.~(\ref{eq:fg}), and the dashed line shows the Feynman-expansion result, Eq.~(\ref{eq:delogi2}), of Ref.~\cite{DeLogi:1977qe}.
 }
\end{figure}

Figure~\ref{Fig:fp_fm_Conv_Exact_Approx_Diff_Q} shows helicity-preserving and helicity-reversing conversion cross sections at low frequencies. The cross sections computed via the partial-wave series in Eq.~(\ref{eq:fg}) [solid] are compared with the approximation obtained via Feynman-diagram expansions [dashed]. More precisely, we compare our numerical results with the cross sections in Eq.~(3.19) of De Logi and Mikelson \cite{DeLogi:1977qe},
\begin{subequations} \label{eq:delogi2}
\begin{align}
\left| \mathfrak{f}_{0} \right|^2 = \frac{1}{4} Q^2 \cot^2 (\theta/2) \left(1 + \cos \theta \right)^2 , \\
\left| \mathfrak{g}_{0} \right|^2 = \frac{1}{4} Q^2 \cot^2 (\theta/2) \left(1 - \cos \theta \right)^2 ,
\end{align}
\end{subequations}
The sum of these terms yields Eq.~(\ref{eq:delogi}), after restoring dimensionful constants. % ($G$, $c$, and $4 \pi \epsilon_0$).

In Fig.~\ref{Fig:fp_fm_Conv_Exact_Approx_Diff_Q}, the comparison is made in the low frequency regime ($2M\omega =0.1$) for three charge-to-mass ratios: $Q=0.01M$, $0.5M$ and $0.8M$. In each case, the conversion cross sections are well described by the approximation in Eq.~(\ref{eq:delogi2}). Consistent closed-form results were also obtained via the Born approximation in Ref.~\cite{Breuer:1981kd}, and again by Feynman-diagram techniques in Ref.~\cite{Bjerrum-Bohr:2014lea}.

\begin{figure}%[htb]
 \includegraphics[scale=0.550]{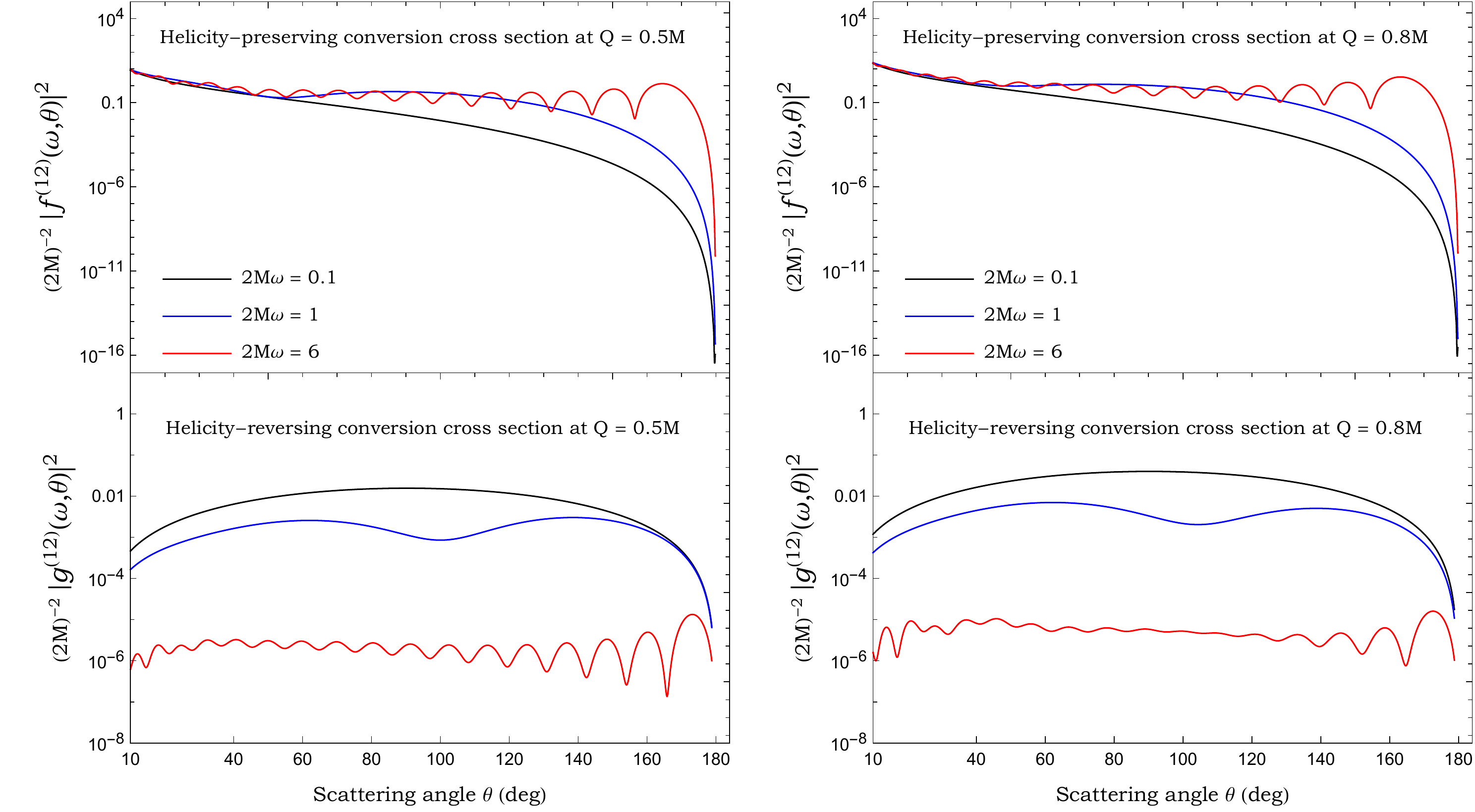}
 \caption{\label{Fig:fp_fm_Conv_Exact_Q_05M_08M_Diff_2Mw} Helicity-preserving and helicity-reversing conversion cross amplitude for $Q=0.5M$ (left panel) and $Q=0.8M$ (right panel) at higher frequencies ($2M\omega = 0.1, 1$ and $6$).}
\end{figure}

At higher frequencies, the conversion cross section develops additional structure.
Figure~\ref{Fig:fp_fm_Conv_Exact_Q_05M_08M_Diff_2Mw} shows the conversion cross section at higher  frequencies ($2M\omega = 0.1$, $1$ and $6$) at $Q = 0.5 M$ and $0.8 M$. The dominant contribution is from the helicity-preserving amplitude, and the helicity-reversing amplitude diminishes as $M \omega$ increases. Relatedly, the phase difference between the odd and even parity modes at fixed $(\ell + 1/2) / \omega$, given in Eq.~(\ref{Rel_2_Ap}), diminishes as $\omega$ increases. In other words, a circularly-polarized incident EW (GW) generates an elliptically-polarized GW (EW) in general; but the elliptical polarization becomes essentially circular at high frequencies.

For $M \omega \gtrsim 1$, Fig.~\ref{Fig:fp_fm_Conv_Exact_Q_05M_08M_Diff_2Mw} shows regular spiral-scattering (`orbiting') oscillations in the cross sections. In the semi-classical interpretation, these are due to the interference between rays scattered by angles $\theta$, $2 \pi - \theta$, $2 \pi + \theta$, $4 \pi - \theta$, etc. (see Fig.~\ref{Fig:ray-plot} and rays (b) and (c)). The relative phase difference associated with the first and second ray is determined by their path difference; as this increases in a nearly linear fashion with $\theta$, the oscillations are regular. Making the crude-but-effective approximation that the rays circulate on the photon orbit yields an approximate angular width of $\pi / (\omega b_c)$, with $b_c$ given in Eq.~(\ref{eq:bc}).

\begin{figure}%[htb]
 \includegraphics[scale=0.57]{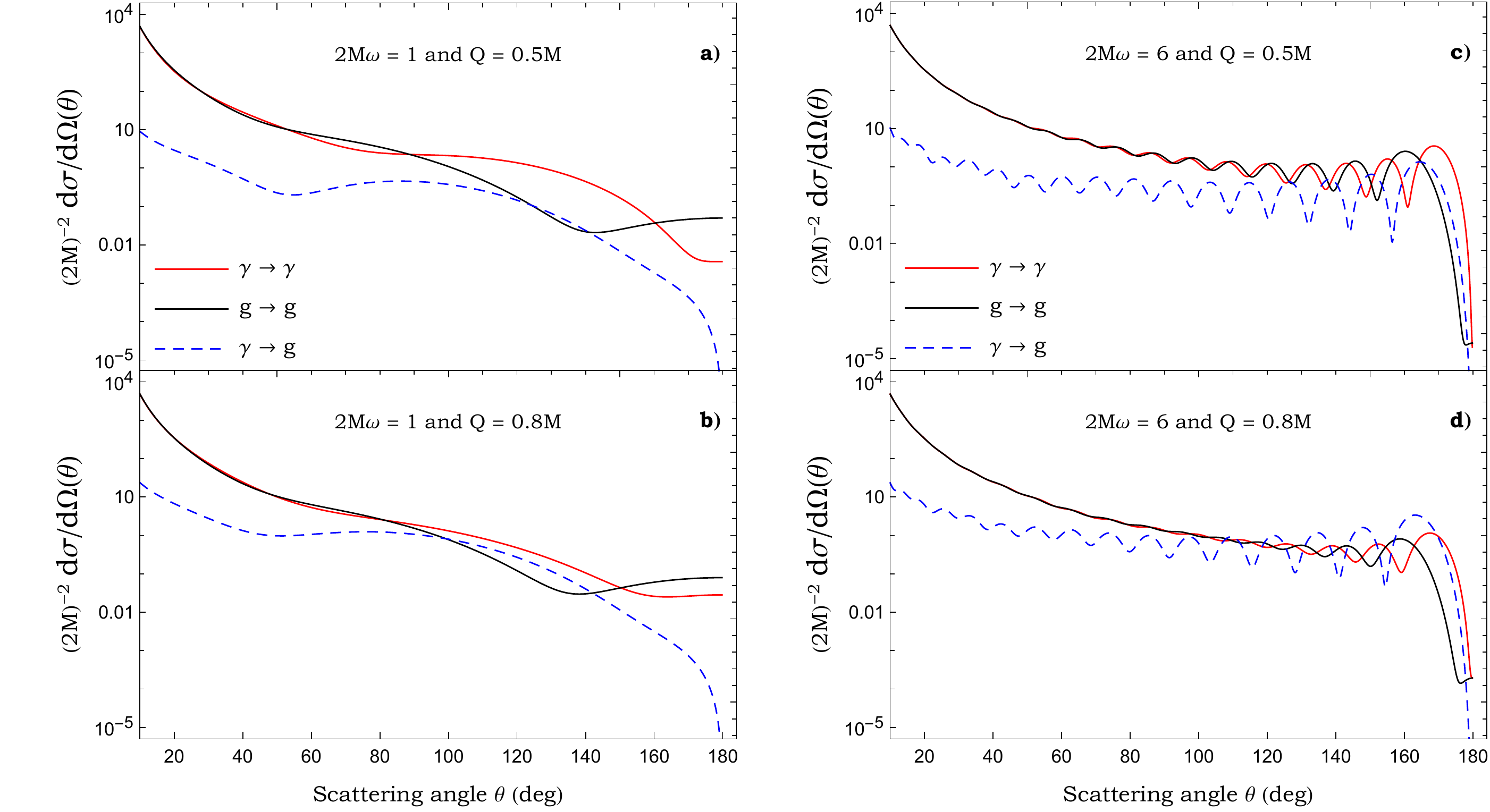}
 \caption{\label{Fig:EMtoEM_vs_GWtoGW_vs_EMtoGW} Scattering and conversion cross sections on the Reissner-Nordstr\"om black hole spacetime. The plots compare the scattering cross sections, labelled $\gamma \rightarrow \gamma$ in the electromagnetic-wave case and $g \rightarrow g$ in the gravitational-wave case, with the conversion cross section labelled $\gamma \leftrightarrow g$. The cross section for conversion of an electromagnetic wave to a gravitational wave is equal to the cross section for the conversion of a gravitational wave to an electromagnetic wave. \emph{Left}: Lower frequency $2M\omega=1$ for (a) $Q =0.5M$ and (b) $2M\omega=1$ and $Q =0.8M$. \emph{Right:} Higher frequency $2M\omega=6$ for (c) $Q =0.5M$ and (d) $Q =0.8M$.}
\end{figure}

Figure~\ref{Fig:EMtoEM_vs_GWtoGW_vs_EMtoGW} compares the conversion cross section ($\gamma \leftrightarrow g$) with the scattering cross sections ($\gamma \to \gamma $ and $g \to g$).
At small angles, the conversion cross section exhibits a $\theta^{-2}$ divergence in the forward direction, whereas the scattering cross sections exhibit a $\theta^{-4}$ divergence, as anticipated in Eq.~(\ref{eq:geo-csec-divergence}). The conversion cross section is exactly zero in the backward direction $\theta = \pi$, but the scattering cross sections are not zero, due to the non-zero amplitudes $\mathfrak{g}^{(22)}$ (for $Q \ge 0$) and $\mathfrak{g}^{(11)}$ (for $Q > 0$ only) \cite{Crispino:2014eea}. At large angles, regular spiral-scattering oscillations are present for $M\omega \gtrsim 1$. Notably, the conversion cross section can actually exceed the scattering cross section at large scattering angles, as is evident in Fig.~\ref{Fig:EMtoEM_vs_GWtoGW_vs_EMtoGW}(d).

%%%%%Comparison EMtoEM vs GWtoGW vs EMtoGW%%%%%%%%%%%%%%%%%%%%%%%

\begin{figure}%[htb]
 \includegraphics[scale=0.670]{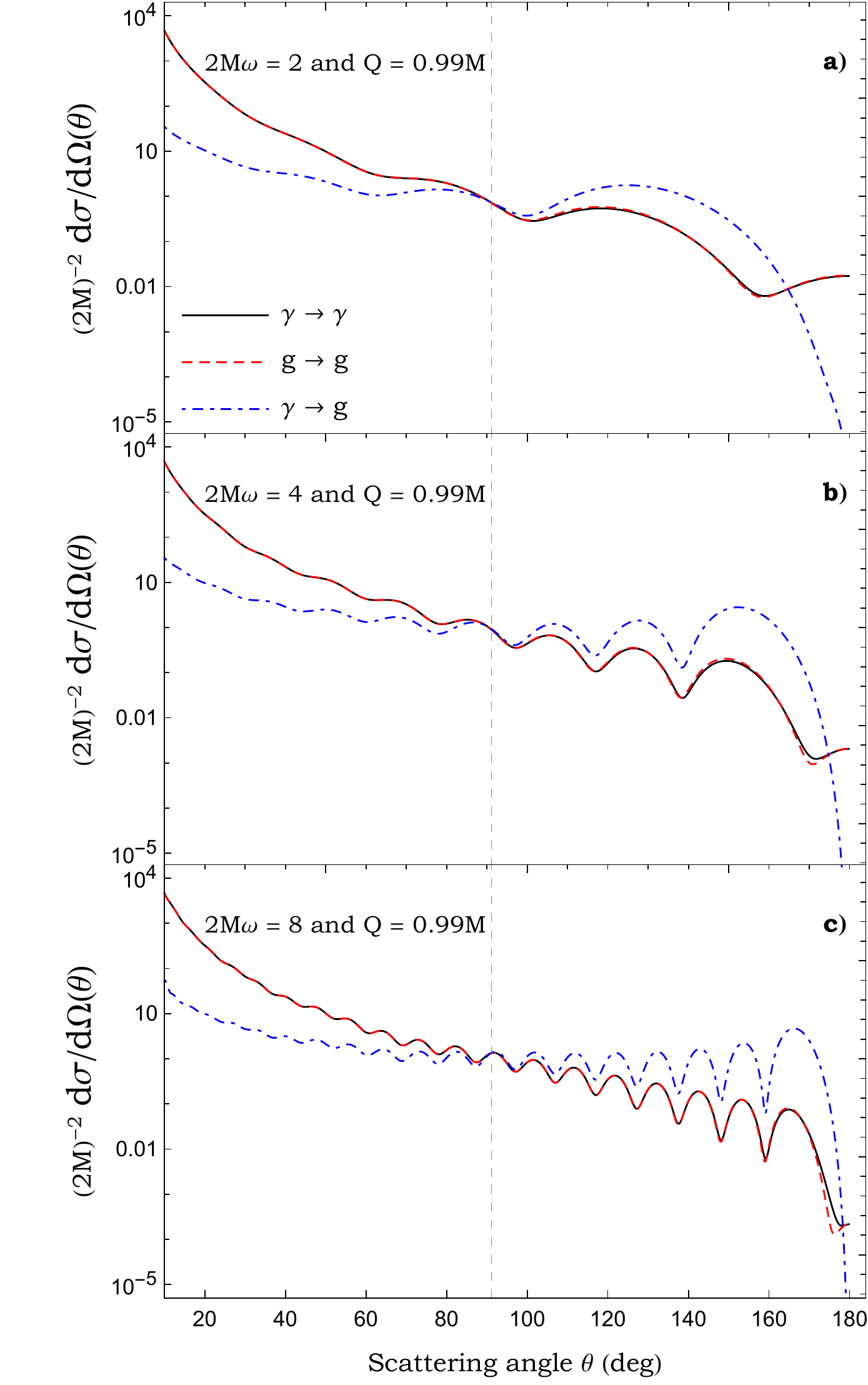}
 \caption{\label{Fig:Conv_Cross_section_Extreme_Q}
  Scattering and conversion cross sections for a near-extremal Reissner-Nordstr\"om black hole ($Q=0.99M$) at three frequencies. The $\gamma \rightarrow \gamma$ and $g\rightarrow g$ scattering cross sections [black and red] are almost equal (see Ref.~\cite{Crispino:2015gua}). The $\gamma \leftrightarrow g$ conversion cross section [blue dot-dashed] exceeds the scattering cross section for angles $\theta \gtrsim 91^\circ$.
 }
\end{figure}

Figure \ref{Fig:Conv_Cross_section_Extreme_Q} shows the scattering and conversion cross sections for a nearly-extremal black hole, with a charge-to-mass ratio $Q/M = 0.99$. At small angles, the scattering cross section ($\sim \theta^{-4}$) dominates over the conversion cross section ($\sim \theta^{-2}$). However, for larger angles $\theta \gtrsim 91^\circ$, the conversion cross section is \emph{larger} than the scattering cross section. In other words, for an incident electromagnetic wave, the energy flux in gravitational waves will exceed that in electromagnetic waves at large angles (and vice versa, for an incident gravitational wave). The three plots show that this effect is rather insensitive to the wave frequency. A satisfying physical explanation for this universality comes from the geometric optics approach devised by Gerlach \cite{Gerlach:1974zz}, which associates a conversion factor with each ray (see Fig.~\ref{Fig:ray-plot} and Sec.~\ref{sec:geo-optics}).

\begin{figure}%[htb]
 \includegraphics[scale=0.97]{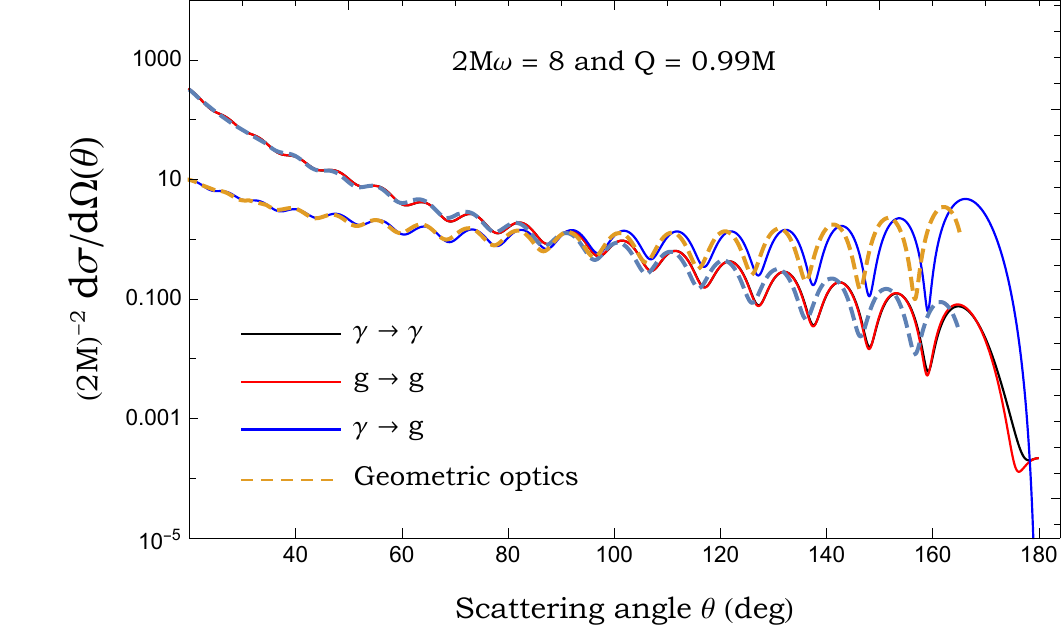}
 \caption{\label{Fig:geometric-optics-comparison}
 The geometric-optics approximation. The solid lines show the scattering and conversion cross sections calculated via the partial-wave expansion. The dashed lines show the geometric-optics approximation of Eqs.~(\ref{eq:geo-ampl})--(\ref{eq:geo-csec}), obtained by solving transport equations along null rays.
  }
\end{figure}

Figure \ref{Fig:geometric-optics-comparison} compares the geometric-optics approximation (Sec.~\ref{sec:geo-optics}) in Eqs.~(\ref{eq:geo-ampl})--(\ref{eq:geo-csec}) [dashed] with the partial-wave cross sections [solid], showing a good qualitative agreement. In the geometric-optics approach, the orbiting oscillations arise from interference between the primary ray, scatted by an angle $\theta$, and the secondary ray, scattered by an angle $2 \pi - \theta$ (see Fig.~\ref{Fig:ray-plot}). The conversion cross section arises from the accumulated conversion phase $\chi$ along a ray. At $\theta \approx 91^\circ$, the conversion phase along the primary ray is $\chi = \pi/4$, implying that half of the energy in the incident wave has been converted (see Eq.~(\ref{eq:conv-oscillation})). It is at this angle that we see the conversion cross section become greater than the scattering cross sections.

\begin{figure}%[htb]
 \includegraphics[scale=1]{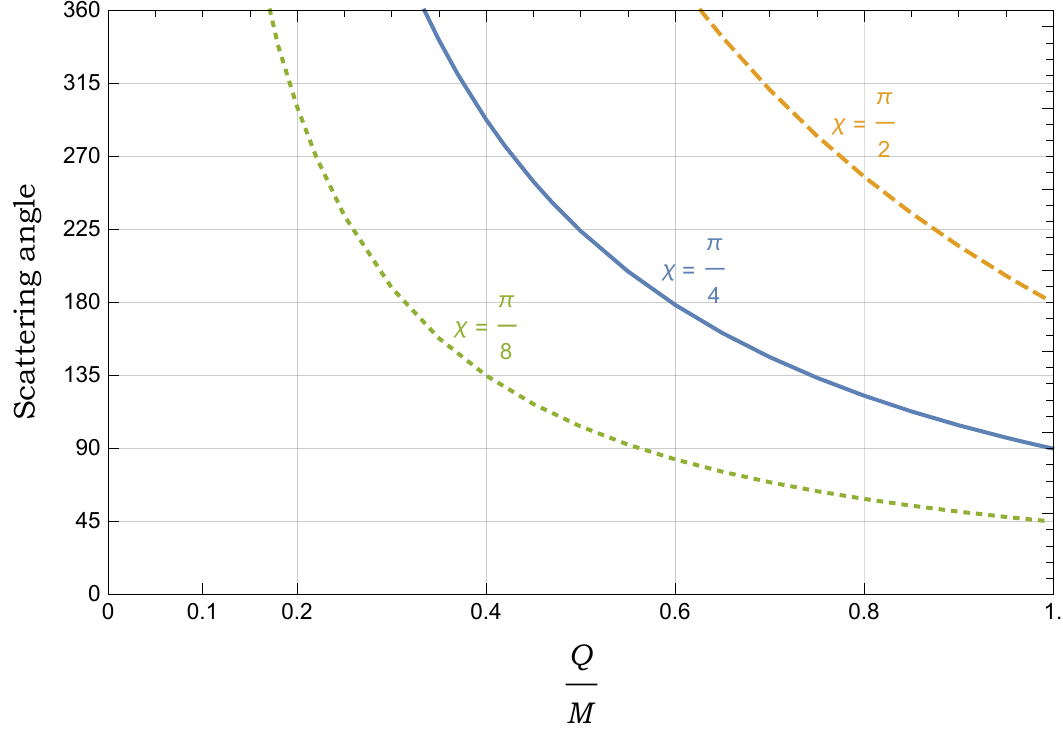}
 \caption{\label{Fig:geometric-optics-rays}
 The scattering angle of the ray corresponding to half-conversion ($\chi = \pi/4$) and total-conversion ($\chi = \pi/2$) of  electromagnetic waves into gravitational waves (and vice versa) in the short-wavelength limit ($M\omega \gg 1$), as a function of charge-to-mass ratio. The curves shown are contours of the conversion phase $\chi$ that arises in the geometric-optics approximation (see Sec.~\ref{sec:geo-optics}).
  }
\end{figure}

Having established the validity of the geometric optics approximation (in the regime $M \omega \gg 1$ and away from the poles), we can now compute the scattering angle for which scattering and conversion are equal (in the short-wavelength limit) by examining the conversion phase $\chi$ in more detail. This angle was calculated numerically via the method of Sec.~\ref{sec:geo-optics}. Figure \ref{Fig:geometric-optics-rays} shows the angle at which $\chi = \pi / 4$ [solid line], as a function of black hole charge, corresponding to half-conversion. The line for total conversion ($\chi = \pi / 2$) is also shown [dashed].

In the extremal limit $Q = M$, the numerical data supports the simple linear relationship between the conversion phase $\chi$ and the deflection angle $\Theta$ derived in Eq.~(\ref{eq:chi-linear}). The implication is that, for an extremal black hole, the converted energy will exceed the scattered energy for angles greater than $90^\circ$, in the high-frequency regime to which the geometric-optics approximation applies. (A caveat here is that the  geometric optics approximation we have used in Sec.~\ref{sec:geo-optics} breaks down for angles close to $\theta = 180^\circ$, as it is no longer valid to consider pairs of distinct rays, but instead a one-parameter family).

\section{Discussion and conclusion\label{sec:conclusions}}

In this work, we have calculated the scattering and conversion cross sections for planar electromagnetic and gravitational waves impinging upon a charged black hole. Accurate numerical results were obtained by summing partial-wave series for the amplitudes, summarized in Sec.~\ref{sec:summary}. In the long-wavelength regime ($M\omega \ll 1$), we found that the conversion cross section matches the Feynman-expansion result (see Eq.~(\ref{eq:delogi}), Eq.~(\ref{eq:delogi2}) and Fig.~\ref{Fig:fp_fm_Conv_Exact_Approx_Diff_Q}). In the short-wavelength regime ($M \omega \gg 1$), the scattering and conversion cross sections are well described by a (numerically-calculated) geometric-optics approximation developed in Sec.~\ref{sec:geo-optics}. This approximation involves calculating Gerlach's conversion phase $\chi$ (introduced in Ref.~\cite{Gerlach:1974zz}) along the primary and secondary null rays (see Fig.~\ref{Fig:geometric-optics-comparison}).

A key finding of this work is that the converted flux can exceed the scattered flux at large angles, if the black hole is sufficiently charged. In other words, the conversion of electromagnetic waves to gravitational waves with the same frequency -- and vice versa -- is substantial for parts of the wavefront that pass close to the circular photon orbit of a highly-charged black hole. Figure \ref{Fig:Conv_Cross_section_Extreme_Q} shows this phenomenon for the case $Q = 0.99M$. In the short-wavelength regime, the (numerical) geometric-optics analysis implies that the converted flux can exceed the scattered flux at large angles if $Q \gtrsim 0.6M$. The scattering angle beyond which the converted flux exceeds the scattered flux is shown in Fig.~\ref{Fig:geometric-optics-rays} [blue solid line]; it reaches a minimum of $90^\circ$ in the extremal case ($Q = M$).

If the incident wave is circular-polarized, the outgoing scattered and converted waves are elliptically-polarized, in general. At low frequency, the effect is encapsulated by the non-zero helicity-reversing conversion amplitude, $\mathfrak{g}_0$ in Eq.~(\ref{eq:delogi2}) \cite{DeLogi:1977qe, Bjerrum-Bohr:2014lea}.  In the partial wave expansion, elliptical polarization arises from the difference in phase shifts between odd and even-parity perturbations (see Eq.~(\ref{eq:g})).  We have demonstrated here that the helicity-reversing amplitudes diminish rapidly as $M\omega$ increases (see Fig.~\ref{Fig:fp_fm_Conv_Exact_Q_05M_08M_Diff_2Mw}), such that in the short-wavelength limit the scattered and converted flux is essentially circular-polarized, with the same handedness as the incident wave.

Here we have investigated gravitoelectrical conversion in the idealised setting of a charged black hole in vacuum. Astrophysical black holes are not expected to sustain sizable charge-to-mass ratios $Q/M$ \cite{Gibbons:1975kk}. Conversion in the more realistic setting of a black hole magnetosphere coupled to an accretion disk was recently examined in Ref.~\cite{Saito:2021}.

An obvious extension of this work is to the \emph{rotating} charged scenario: the Kerr-Newman black hole. The scattering of a scalar field by a Kerr-Newman black hole was studied in \cite{Leite:2019eis}. For electromagnetic and gravitational perturbations, a separation of variables has not been achieved, and alternative approaches \cite{Pani:2013ija} are necessary to solve the coupled equations in full. On the other hand, the geometric-optics approximation should be straightforward, as the geodesic equations for null rays are (Liouville) integrable.

\begin{acknowledgments}
S.D.~thanks Devin Walker, Vitor Cardoso, and Panagiotis Giannadakis for discussions. 
S.D.~acknowledges financial support from the Science and Technology Facilities Council (STFC) under Grant No.~ST/P000800/1, and from the European Union's Horizon 2020 research and innovation programme under the H2020-MSCA-RISE-2017 Grant No.~FunFiCO-777740.  
\end{acknowledgments}

\bibliography{rn_em_gw}

\end{document}